\documentclass[aps,prl,twocolumn,letterpaper,superscriptaddress,amsmath,citeautoscript]{revtex4-1}

\usepackage{graphicx}
\usepackage{mathcomp,fixmath}

\usepackage[TS1,T1]{fontenc}
\usepackage{ucs}\usepackage[utf8x]{inputenc}
\usepackage[british]{babel}
\usepackage[draft=false]{microtype}

\usepackage{array,dcolumn,ltxtable,booktabs}
\usepackage[colorlinks,citecolor=blue,filecolor=blue,linkcolor=blue,urlcolor=blue]{hyperref}

\renewcommand{\mathbf}[1]{\boldsymbol{#1}}

\begin{document}

\title{Transport of Intensity Phase Retrieval of Arbitrary Wave Fields Including Vortices}

\author{Axel Lubk}
\email{axel.lubk@triebenberg.de}
\affiliation{Speziallabor Triebenberg, Technische Universit\"{a}t Dresden, 01062 Dresden, Germany}
\affiliation{EMAT, Universiteit Antwerpen, Groenenborgerlaan 171, 2020 Antwerpen, Belgium}
\author{Giulio Guzzinati}
\affiliation{EMAT, Universiteit Antwerpen, Groenenborgerlaan 171, 2020 Antwerpen, Belgium}
\author{Felix Börrnert}
\affiliation{Speziallabor Triebenberg, Technische Universit\"{a}t Dresden, 01062 Dresden, Germany}
\author{Jo Verbeeck}
\affiliation{EMAT, Universiteit Antwerpen, Groenenborgerlaan 171, 2020 Antwerpen, Belgium}

\begin{abstract}
The phase problem can be considered as one of the cornerstones of quantum mechanics intimately connected to the detection process and the uncertainty relation. The latter impose fundamental limits on the manifold phase reconstruction schemes invented to date in particular at small magnitudes of the quantum wave. Here, we show that a rigorous solution of the Transport of Intensity Reconstruction (TIE) scheme in terms of a linear elliptic partial differential equation for the phase provides reconstructions even in the presence of wave zeros if particular boundary conditions (BCs) are given. We furthermore discuss how partial coherence hampers phase reconstruction and show that a modified version of the TIE reconstructs the curl-free current density at arbitrary (in-)coherence. This opens the way for a large variety of new applications in fields as diverse as astrophysics, geophysics, photonics, acoustics, and electron microscopy, where zeros in the respective wave field are a ubiquiteous feature.
\end{abstract}

\maketitle

Ever since the introduction of quantum mechanical wave equations the loss of phase information in the detection process of particles has stirred scientists to the invention of numerous methods to retrieve the missing information. Most notably, a variety of holographic schemes \cite{Cowley(1992)} based on coherently superimposing known reference waves to the wave field has been applied successfully to recover phases of matter and photon waves \cite{Gabor(1948)a}. One particular holographic scheme is referred to as Transport of Intensity Equation (TIE) reconstruction \cite{Teague(1983)}. Because of its simple and flexible experimental setup TIE phase retrievals have been reported for waves consisting of atoms \cite{Fox(2002)}, neutrons \cite{Allman(2000)}, X-rays \cite{Nugent(2010)}, electrons \cite{Ishizuka(June2005)}, and visible light \cite{Barty(1998)}. Similar to all holographic schemes, two fundamental limits prevail: First, partial coherence obscures the meaning of reconstructed phases \cite{Gureyev(
1995),Zysk(2010)} and second, due to density-phase uncertainty relations holographic reconstructions at $\rho\ll1$ loci suffer from increased phase noise \cite{Fick(1988),*Lenz(1988)}. The TIE method can be considered as the infinitesimal version of Gabor's original inline holography \cite{Beleggia(2004)} and therefore has the advantage to not rely on off-axis reference waves. It is furthermore a linear reconstruction scheme and peculiar in that a differential equation is involved. The TIE scheme is based on the equation of continuity \cite{Teague(1983)} 
\begin{eqnarray}
\frac{\partial\rho\left(\mathbf{r},z\right)}{\partial z} & = & -\frac{1}{k}\nabla\cdot\mathbf{j}\left(\mathbf{r},z\right)\label{eq:TIE}\\
 & = & -\frac{1}{k}\nabla\cdot\left(\rho\left(\mathbf{r},z\right)\nabla\varphi\left(\mathbf{r},z\right)\right)\,,\nonumber 
\end{eqnarray}
derived from the stationary paraxial wave equation 
\begin{equation*}
\frac{\partial\Psi\left(\mathbf{r},z\right)}{\partial z}=\frac{i}{2k}\triangle\Psi\left(\mathbf{r},z\right)
\end{equation*}
valid for a large variety of scattering phenomena of, e.\,g.\ electrons, photons, or atoms,  moving within a small solid angle around the $z$-axis \cite{Schmalz(2011)}. Here, $k$ is the wave number, $\mathbf{r}=\left(x,y\right)^{T}$ the 2D position vector, $\nabla=\left(\partial_{x},\partial_{y}\right)^{T}$ the 2D gradient, $\mathbf{j}$ the 2D current density vector, and $\rho=\left|\Psi\right|^{2}$ the particle density. In order to recover the phase $\varphi$ one records at least two slightly defocused images\,---\,$\rho\left(z-\delta z\right)$ and $\rho\left(z+\delta z\right)$ where the wave optical defocus corresponds to propagation along $z$\,---\,and approximates
\begin{subequations}
\label{eq:drhodzrho}
\begin{equation}
\frac{\partial\rho\left(z\right)}{\partial z}=\frac{\rho\left(z+\delta z\right)-\rho\left(z-\delta z\right)}{2\delta z}+\mathcal{O}\left(\delta z^{2}\right)\label{eq:drhodz}
\end{equation}
and
\begin{equation}
\rho\left(z\right)=\frac{\rho\left(z+\delta z\right)+\rho\left(z-\delta z\right)}{2}+\mathcal{O}\left(\delta z^{2}\right)\label{eq:rho}
\end{equation}
\end{subequations}
in Eq.\:\eqref{eq:TIE}. It has been noted that the $\mathcal{O}\left(\delta z^{2}\right)$ errors to the approximations \eqref{eq:drhodzrho} remain small if the Fresnel parameter $N_{F}=h^{2}\cdot\left(k/\delta z\right)\gg 1$ for a typical object feature~$h$ \cite{Dyck(1987),Beleggia(2004),Gureyev(2004)}. However, we show below that the influence of the errors on the reconstructed phase can still grow large.

Previously, for solving the TIE one assumed either Dirichlet BCs where the phase at the boundary is set to a fixed value, mostly $\varphi\left(\infty\right)=0$ \cite{Gureyev(2003)} or periodic BCs, reducing Eq.\:(\ref{eq:TIE}) to an algebraic problem in Fourier space \cite{Dyck(1987),Gureyev(1998),Paganin(1998),Ishizuka(June2005)}. Most of the solutions were furthermore based on the rather restrictive assumptions of a pure phase object $\rho=\mathrm{const.}$ \cite{Gureyev(1998)} or a conservative current density, where $\mathbf{j}=\nabla\chi$ is the gradient of a scalar potential $\chi$ and Eq.\:\eqref{eq:TIE} reduces to a Poisson equation \cite{Teague(1983),Paganin(1998),Allen(2001),Ishizuka(June2005)}. Most notably, $\rho=0$ loci were either completely excluded\cite{Dyck(1987),Gureyev(1998),Ishizuka(June2005),Schmalz(2011)}, stated to yield non-unique results \cite{Gureyev(1995),Peele(2004)}, or discussed by Helmholtz decomposing $\nabla\varphi$ \cite{Paganin(1998)a,*Nugent(2000),*Allen2(2001)}. While the 
first two statements are correct within their context, the latter has to be treated with caution because $\nabla\varphi$ remains undefined at $\rho=0$.

Zeros in the wave field such as phase singularities (vortices) are a general feature \cite{Nye(1974)}, for instance, a complicated pattern of phase vortices already occurs upon interference of only three plane waves \cite{O'Holleran(2006)}. In particular, optically created photonic and electron vortex beams attracted considerable attention for their interesting physical properties and potential applications \cite{Molina-Terriza(2007),*Franke-Arnold(2008),*Lloyd(2012),Verbeeck(2010)}. We discuss the details of the TIE's experimental implementation with a focus on electron waves occurring in the transmission electron microscope (TEM) and reconstruct the phase of an experimental electron vortex beam (and of a complicated numeric test wave, see Supplementary Information) as a proof-of-principle. Also, we discuss implications arising from partial coherence and experimental noise without touching the field of possible regularization schemes to the latter problem.

First, we consider the TIE as an elliptic problem. Equation~\eqref{eq:TIE} with a given density $\rho$ and density derivative $\partial\rho/\partial z$ is an inhomogeneous linear elliptic PDE for the phase $\varphi$. The corresponding theory based on the Lax-Milgram theorem ensures existence and uniqueness of weak solutions within \emph{simply connected domains} with $\rho>0$ and mixed Dirichlet, von\:Neumann (fixed derivative $\nabla_{\mathbf{n}}\varphi$ normal to the boundary), or periodic BCs \cite{Gilbarg(1983),*Evans(1998)}. The latter two BCs do not fix a constant phase offset. The substitution $\rho\nabla\varphi=\nabla\chi$ \cite{Teague(1983)}, transforming \eqref{eq:TIE} into a Poisson problem for $\chi$, is only valid if the 2D rotation or vorticity of the current density, denoted by the wedge product $\nabla\wedge \mathbf{j}=\left(\partial_{x}j_{y},-\partial_{y}j_{x}\right)$, vanishes $\nabla\wedge\left(\rho\nabla\varphi\right)=0$ \footnote{$\nabla\wedge\left(\rho\nabla\varphi\right)=\nabla\Psi^{*}\
wedge\nabla\Psi=\nabla\rho\wedge\nabla\varphi+\rho\nabla\wedge\nabla\varphi=0$} \cite{Frankel(1999)}. This particularly excludes phase singularities $\nabla\wedge\nabla\varphi\neq0$ \footnote{Note that expressions involving the phase at $\rho=0$ are only defined in a distributional sense.}. In contrast to $\varphi$, the current $\mathbf{j}$ is well-defined everywhere\,---\,even at $\rho=0$\,---\,implying that a 2D Helmholtz decomposition $\mathbf{j}=\nabla\alpha+(1,1)^{T}\wedge\nabla\beta$ with some scalar functions $\alpha$, $\beta$ is well-defined \footnote{In 2D the divergence free current is merely a 90$\text{\textdegree}$ rotated curl free one.}. When inserting this decomposition into the TIE we obtain
\begin{eqnarray}
\frac{\partial\rho\left(\mathbf{r},z\right)}{\partial z} & = & -\frac{1}{k}\nabla\cdot\left(\nabla\alpha\left(\mathbf{r},z\right)+(1,1)^{T}\wedge\nabla\beta\left(\mathbf{r},z\right)\right)\\
 & = & -\frac{1}{k}\triangle\alpha\left(\mathbf{r},z\right)\,.\nonumber 
\end{eqnarray}
Consequently, the Poisson formulation of the TIE can be used to reconstruct the curl-free current density $\mathbf{j}_{\alpha}=\nabla\alpha$ instead of the phase. In contrast to the phase the current density is exactly defined even for arbitrary mixed states, that is, even in the presence of partial coherence. This can be useful, for instance, to find vortices in a partially coherent wave field.

\begin{figure}
\includegraphics{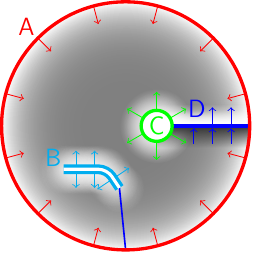}
\begin{tabular}{ccc}
\toprule 
BC type & $\varphi_{l}$ & $\nabla_{\mathbf{n}}\varphi_{l}$\\
\midrule
A &  & 0\\
B & $m\,\pi+\varphi_{r}$ & $\nabla_{\mathbf{n}}\varphi_{r}$\\
C &  & 0\\
D & $w\,2\pi+\varphi_{r}$ & $\nabla_{\mathbf{n}}\varphi_{r}$\\
\bottomrule
\end{tabular}
\caption{Scheme of boundary condition (BC) types with arrows indicating the defined directional derivatives. A\,---\,outer von\:Neumann BC, B\,---\,normal zero with constant derivative and possible $m=0,\pm1$ phase jump, C\,---\,von\:Neumann BC around a vortex, and D\,---\,phase sheet change with constant derivative and $w\,2\pi$ phase jump. $\varphi_{l,r}$ denote the phases on the left/right hand side of the boundary (the latter is only defined at interior boundaries B and D).
\label{fig:Scheme-of-possible}}
\end{figure}
The TIE \eqref{eq:TIE} knows nothing about multi-valued or undefined phases. Indeed, the latter has to be taken into account by providing the appropriate BCs for the phase at loci where $\rho=0$ or phase sheet changes to the adjacent $2\pi$ phase interval occur. Figure~\ref{fig:Scheme-of-possible} outlines possible BC types. In addition to the outer BCs denoted with A, three different inner BCs have to be considered therefore: B at manifolds of normal zeros of $\rho$ without phase singularities, C at phase singularities with winding number $w=1/2\pi\oint\mathrm{d}\mathbf{s}\cdot\nabla\varphi\neq0$ around the singularity, and D at lines where the phase passes to the next phase sheet. While the phase jumps $m=\left(0,\pm1\right)\cdot\pi$ at B and $w\cdot2\pi$ at D are obvious, the BCs at A and C of the von\:Neumann type are chosen in the following because unlike the other BC types they do not restrict the wave topology. Furthermore, cuts have to be introduced from each isolated zero to the boundary in order to 
render the domain simply connected again. Here, the BC at the cut determine the topology of the zero. 

In order to facilitate an independent choice of the winding numbers $w_{n}$ around all $N$ singularities in the wave field one therefore has to define $N$ lines starting at $N$ vortices and ending at the outer boundary. As an analogon, we note that similar BCs occur for the displacement field around dislocations in solids and are referred to as Volterra construction in that context \cite{Hirth(1967)}. Below, we therefore adapt numerical schemes from the elastic theory based on the Finite Element Analysis to solve the TIE \cite{Gracie(2007)}. We also note that simple zeros and vortices might superimpose in arbitrary ways.

The central obstacle towards TIE phase reconstruction is now finding the correct BCs to the unknown phase. Our solution to that problem is based on three observations: (i) As a consequence of the linear nature of the TIE, solutions to close-to-the-exact BCs are close to the exact solution. This is the reason why approximate von\:Neumann BCs for type A and C are acceptable in the examples that follow. The remaining degree of freedom of the BCs is now reduced to all possible combinations of phase jumps $m_{k}$ for a given set of $K$ normal zero manifolds and winding numbers $w_{n}$ for a given set of $N$ singularities and cuts connecting them to the outer boundary. (ii) Although different BCs yield acceptable solutions to Eq.\:\eqref{eq:TIE} they show a different behaviour at large defoci. The latter facilitates a consistency check with a reference density $\rho\left(z+a\delta z\right)$, $\left|a\right|\gg1$ singling out the correct solution. In practice, one would start to test the small $m_{k}$ and $w_{n}$ 
first, which usually leads to the solution rather quickly, particularly if taking into account that higher order vortices are usually unstable and dissolve into first order ones \cite{Freund(1999),Lubk(2013)}. (iii) Since TIE solutions can be equivalently obtained at different defoci, $z$ can be tuned to reduce the set of normal zeros and consequently the combinatorial problem. A larger set of differently defocused images also helps in finding phase singularities because\,---\,in contrast to normal zeros\,---\,they are topologically protected and thus cannot be destroyed. Consequently, vortices show up as stable minima of both the density and the reconstructed curl-free current in the defocus series and can be identified accordingly.

In the following, we discuss an experimental adaptation of the outlined scheme to electron wave reconstruction using an electron vortex beam as example. The electron vortex beam was generated by inserting a Fresnel zone plate containing an edge dislocation (``fork aperture'') into the condenser aperture of a non-hardware-aberration-corrected Philips CM30 TEM \cite{Verbeeck(2010)}. The beam was focused in the sample plane with a semi convergence angle of $0.36\pm0.02$\:mrad and the defocus series encompassed $-17$\:\textmu m to $17$\:\textmu m with a $1$\:\textmu m step size (see Supplementary Information). 

\begin{figure*}
\includegraphics[width=\textwidth]{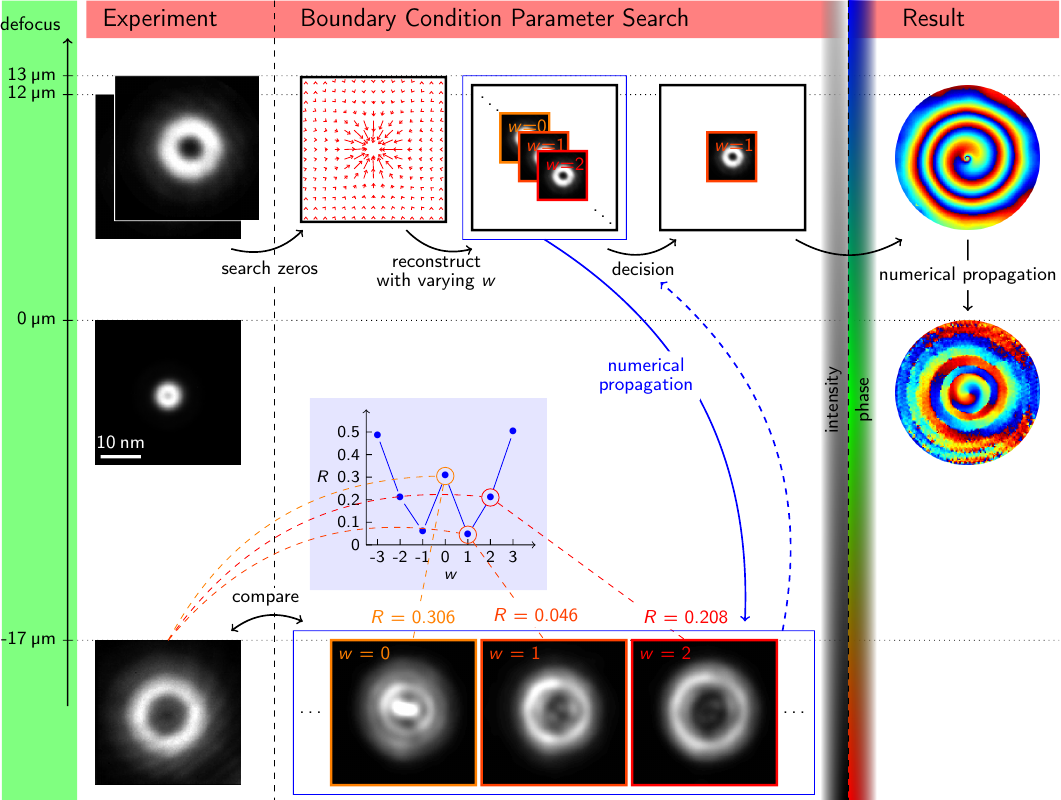}
\caption{TIE reconstruction scheme of a vortex beam. The experiment column on the left shows selected recorded densities from a defocus series. For the BC parameter search, a set of possible defocused solutions to $w=\ldots,0,1,2,\ldots$ vortices is computed and propagated to another defocus. All propagated solutions are compared to the experimental image recorded at this particular defocus value with $w=1$ fitting best according to $R$ factor consistency. The result column on the right shows the reconstructed phase with a $w=1$ singularity and finally the phase of the numerically focused reconstructed wave. The scale bar in the in focus micrograph applies for all images in the scheme.\label{fig:TIE-reconstruction-scheme:}}
\end{figure*}

Figure~\ref{fig:TIE-reconstruction-scheme:} outlines the general workflow of the phase reconstruction. In the first column on the left the recording of the data in the experiment is sketched, in this example it is a set of electron micrographs. The TIE reconstruction is based on evaluating densities recorded at small defocus steps. As a consequence of the discussion above one typically records a whole series of images over a larger range of defoci to have a bigger choice of $z$ planes at hand. While recording such a series the image might fluctuate in overall intensity, shift and rotate. Aligning the images and eventually removing additional distortions requires an additional preprocessing step prior to the actual reconstruction (see Supplementary Information). The actual phase reconstruction starts by finding all zeros in the wave field in order to define the loci where BCs have to be defined. This can be very tricky experimentally because of noise, partial coherence, and sampling, which obscure the 
original zeros in the wave field. The example of the focused vortex beam at $z=0$\:\textmu m depicted in Fig.\:\ref{fig:TIE-reconstruction-scheme:} shows a missing central zero hampering a TIE reconstruction. At larger defoci the central zero is preserved and can be easily detected based on the single stable minimum observable in both curl-free current and density. Thus we used two defocused images at $12$\:\textmu m and $13$\:\textmu m for defining a single 0D central zero.

Equation\:\eqref{eq:TIE} with corresponding BCs can now be solved with the help of Finite Element Analysis similar to the method used by Gracie \emph{et al.} \cite{Gracie(2007)}. This includes generating a simplex mesh on the structured domain containing holes around the 0D phase singularities and the corresponding cuts from the singularities to the outer boundary. The mesh can be adaptively refined in particular close to vortices. Furthermore, $\rho\left(z\right)$ and $\rho\left(z+\delta z\right)$ have to be interpolated on that simplex grid. The solver now generates solutions to a possible set of BCs, in our example $w\in[-3,3]$. Then, each of the set of generated solutions is propagated numerically to a predefined defocus value and compared to an experimental image at this defocus ($z=-17$\:\textmu m in our example) in order to single out the correct solution. Partial coherence is incorporated phenomenologically by convolving the reconstructed densities with the demagnified source size. We quantify the 
deviation from the experimentally defocused image by the $R=\sum\left(\rho_{exp}-\rho_{rec}\right)^{2}/\left(\rho_{exp}\right)^{2}$ factor. Accordingly, $w=1$ fits best although $w=-1$ is also possible and theoretically has the same shape under perfect rotational symmetry. This is consistent with the input due to the ``fork aperture'' only slightly breaking the symmetry. The remaining deviations between the propagated TIE wave and the experimentally defocused image mainly stem from the influence of partial coherence and noise (see below). 

We can now analyze the reconstructed wave $w=1$ at the reconstruction plane $z=12.5$\:\textmu m and the focal plane $z=0$\:\textmu m typically used for application as shown in the result column on the right of figure~\ref{fig:TIE-reconstruction-scheme:}. In the reconstruction plane a defocus producing a spiraling vortex structure in the phase is clearly apparent. In focus the desired annular vortex structure is approximated (see Ref.\:\onlinecite{Lubk(2013)} for analytic expressions) with small deviations being visible, e.\,g.\:due to non-corrected spherical aberration. Note that these deviations can have important consequences on inelastic scattering cross-section, etc.

\begin{figure}
\includegraphics{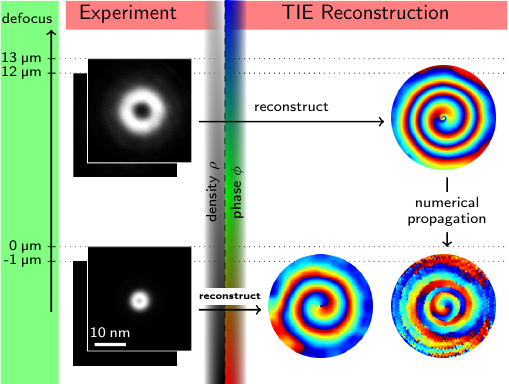}
\caption{Influence of partial coherence on the TIE reconstruction result. In  the experimental column, the starting points for a TIE reconstruction are indicated at in focus and strong defocus. The in focus set is reconstructed directly into a phase map while the defocused set is reconstructed and consequently propagated numerically into focus. The obvious differences between the results are due to the partial coherence violating the reconstruction of the in focus image set. The scale bar in the in focus micrograph applies for all images in the scheme. \label{fig:a)-reconstructed-phase}}
\end{figure}

Finally, we turn to the important discussion of the influence of partial coherence or mixed states, and on the same footing noise and the $\mathcal{O}\left(\delta z^{2}\right)$ correction terms in Eqs.\:\eqref{eq:drhodzrho} on the TIE reconstruction. Strictly speaking, the notion of a single pure state phase is not applicable to mixed states anymore. Instead, phases of mixed states are defined within the context of the particular holographic reconstruction. However, such a phase can differ substantially from the pure state one. For instance, it can be shown that energy-loss off-axis electron holography generally reconstructs complex valued off-diagonals of the density matrix with a phase that does not necessarily correspond to the projected potential obtained in the zero-loss case. In the case of the TIE the phase for a mixed state is \emph{defined} by the TIE and we have to analyze the impact of impurity or noise within this definition. To this end we consider the impact of small incoherent fluctuations $\
delta\rho$ around a pure state $\rho_{p}=\left|\Psi_{p}\right|^{2}$ with corresponding phase $\varphi_{p}$ in the TIE reconstruction 
\begin{equation}
\frac{\partial\left(\rho_{p}+\delta\rho\right)}{\partial z}=-\frac{1}{k}\nabla\cdot\left(\rho_{p}+\delta\rho\right)\nabla\left(\varphi_{p}+\delta\varphi\right)\,.\label{eq:totalTIE}
\end{equation}
Subtracting Eq.\:\eqref{eq:TIE} with $\rho=\rho_{p}$ and $\varphi=\varphi_{p}$ one obtains an elliptic PDE for the phase deviation $\delta\varphi$
\begin{equation}
\frac{\partial\left(\delta\rho\right)}{\partial z}+\frac{1}{k}\nabla\cdot\delta\rho\nabla\varphi_{p}=-\frac{1}{k}\nabla\cdot\rho_{p}\nabla\delta\varphi\label{eq:dphi}
\end{equation}
with the right hand side (homogeneous part) being equivalent to the pure state one described with Eq.\:\eqref{eq:TIE}. The phase deviation is now solved similarly to the TIE, however, with homogeneous Dirichlet BCs at the boundaries and strictly periodic BCs at inner cuts in order to not modify the BCs of the total TIE\:\eqref{eq:totalTIE}. Considering the affine structure of the solutions of an inhomogeneous linear PDE, $\delta\varphi$ can grow large if the inhomogeneous perturbation term surmounts the pure state one, which can easily happen e.\,g.\ at vortices where $\partial\rho_{p}/\partial z\ll1$. Indeed, figure~\ref{fig:a)-reconstructed-phase} shows that for our example the TIE phase reconstructed directly from the in focus images $z=-1$\:\textmu m and $z=0$\:\textmu m significantly deviates from the in focus phase computed \textit{via} the defocused images, confirming the analytic argument. This shows that TIE phase reconstructions from partially coherent mixed states require a certain degree of 
coherence in order to be interpreted as a pure state phase.

In conclusion, we have demonstrated that a rigorous treatment of the Transport of Intensity Equation (TIE) scheme as an elliptic partial differential equation with appropriate boundary conditions (BCs) extends its applicability to arbitrary wave functions containing zeros and singularities. The correct BCs can be found based on topological arguments and a consistency check with reference intensities at larger defoci. We also showed how TIE reconstructed phases are obscured in the presence of partial coherence. The prospects of TIE phase reconstruction are expected to highly benefit from the extended scope: For example, atomic scale electron wave functions are known to contain vortices \cite{Allen3(2001),Lubk(2013)} and applying TIE without taking this into account leads to wrong results. Also in studying vortex beams, this refinement of TIE is crucial as in this case, vortices are present by design. Provided sufficient beam coherence, e.\,g.\ by employing field emission electron sources in TEMs or LASERs in 
photonics, the biggest challenge for the method will be to further refine the BC definition consisting of reliable zero determination and the fast implementation of the combinatorial testing of all possible BCs. Within this context we also mention that accurate TIE solutions can be used as a starting condition for the widely used iterative nonlinear inline reconstruction schemes \cite{Gerchberg(1972)} which then suffer less from stalling and non-uniqueness problems if the iteration starts close to the true solution \cite{Gureyev(2003)}.

\begin{acknowledgements}
The authors acknowledge financial support from the European Union under the Seventh Framework Programme under a contract for an Integrated Infrastructure Initiative. Reference 312483 -- ESTEEM2.
\end{acknowledgements}

\end{document}